\documentclass[twocolumn]{aastex63}

\usepackage[utf8]{inputenc}
\usepackage[english]{babel}
\usepackage{graphicx}

\shorttitle{Near L-edge photoionization of doubly charged iron ions}
\shortauthors{Schippers et al.}

\begin{document}

\title{Near L-Edge Single and Multiple Photoionization of Doubly Charged Iron Ions}

\correspondingauthor{Stefan Schippers}
\email{stefan.schippers@physik.uni-giessen.de}

\author[0000-0002-6166-7138]{Stefan Schippers}
\affiliation{I. Physikalisches Institut, Justus-Liebig-Universit\"{a}t Gie{\ss}en, Heinrich-Buff-Ring 16, 35392 Giessen, Germany}

\author[0000-0001-5100-4229]{Randolf Beerwerth}
\affiliation{Helmholtz-Institut Jena, Fr\"obelstieg 3, D-07743 Jena, Germany}
\affiliation{Theoretisch-Physikalisches Institut, Friedrich-Schiller-Universit\"at Jena, D-07743 Jena, Germany}

\author[0000-0003-3985-2051]{Sadia Bari}
\affiliation{Deutsches Elektronen-Synchrotron DESY, Notkestraße 85, 22607 Hamburg, Germany}

\author[0000-0003-3337-1740]{Ticia Buhr}
\affiliation{I. Physikalisches Institut, Justus-Liebig-Universit\"{a}t Gie{\ss}en, Heinrich-Buff-Ring 16, 35392 Giessen, Germany}

\author[0000-0001-8809-1696]{Kristof Holste}
\affiliation{I. Physikalisches Institut, Justus-Liebig-Universit\"{a}t Gie{\ss}en, Heinrich-Buff-Ring 16, 35392 Giessen, Germany}

\author[0000-0002-8805-8690]{A. L. David Kilcoyne}
\affiliation{Advanced Light Source, MS 80RO114, Lawrence Berkeley National Laboratory, Berkeley, California 94720, USA}

\author[0000-0002-0700-3875]{Alexander Perry-Saßmannshausen}
\affiliation{I. Physikalisches Institut, Justus-Liebig-Universit\"{a}t Gie{\ss}en, Heinrich-Buff-Ring 16, 35392 Giessen, Germany}

\author{Ronald A. Phaneuf}
\affiliation{Department of Physics, University of Nevada, Reno, Nevada 89557-0220, USA}

\author{Simon Reinwardt}
\affiliation{Institut f\"{u}r Experimentalphysik, Universit\"{a}t Hamburg, Luruper Chaussee 149, 22761 Hamburg, Germany}

\author[0000-0002-1111-6610]{Daniel Wolf Savin}
\affiliation{Columbia Astrophysics Laboratory, Columbia University, 550 West 120th Street, New York, New York 10027, USA}

\author{Kaja Schubert}
\affiliation{Deutsches Elektronen-Syhchrotron, DESY, Notkestraße 85, 22607 Hamburg, Germany}

\author[0000-0003-3101-2824]{Stephan Fritzsche}
\affiliation{Helmholtz-Institut Jena, Fr\"obelstieg 3, D-07743 Jena, Germany}
\affiliation{Theoretisch-Physikalisches Institut, Friedrich-Schiller-Universit\"at Jena, D-07743 Jena, Germany}

\author[0000-0002-1228-5029]{Michael Martins}
\affiliation{Institut f\"{u}r Experimentalphysik, Universit\"{a}t Hamburg, Luruper Chaussee 149, 22761 Hamburg, Germany}

\author[0000-0002-0030-6929]{Alfred M\"{u}ller}
\affiliation{Institut f\"{u}r Atom- und Molek\"{u}lphysik, Justus-Liebig-Universit\"{a}t Gie{\ss}en, Leihgesterner Weg 217, 35392 Giessen, Germany}



\begin{abstract}
Using the photon-ion merged-beams technique at a synchrotron light source, we have measured relative cross sections for single and up to five-fold photoionization of Fe$^{2+}$ ions in the energy range 690--920~eV. This range contains thresholds and resonances associated with ionization and excitation of $2p$ and $2s$ electrons. Calculations were performed to simulate the total absorption spectra. The theoretical results show very good agreement with the experimental data, if overall energy shifts of up to 2.5~eV are applied to the calculated resonance positions and assumptions are made about the initial experimental population of the various levels of the Fe$^{2+}$([Ar]$3d^6$) ground configuration. Furthermore, we performed extensive calculations of the Auger cascades that result when an electron is removed from the $2p$ subshell of Fe$^{2+}$. These computations lead  to a  better agreement with the measured product-charge-state distributions as compared to earlier work. We conclude that the $L$-shell absorption features of low-charged iron ions are useful for identifying gas-phase iron in the interstellar medium and for discriminating against the various forms of condensed-phase  iron bound to composite interstellar dust grains.
\end{abstract}

\keywords{atomic data --- atomic processes --- line: identification  --- opacity}



\section{Introduction}

$L$-shell photoionization data of low-charged iron ions are required to reliably assess the abundance of iron in the interstellar medium (ISM) from astronomical x-ray observations. Resonant $L$-shell absorption features offer the possibility to discriminate between iron in the gas phase and iron that is chemically bound in dust grains. Absorption data for chemically bound iron are available from the literature \citep[e.g.,][]{Kortright2000,Lee2009,Miedema2013,Westphal2019}. However, corresponding experimental data for low-charged iron ions had been missing. In order to fill this gap we have launched a campaign to provide absorption data for Fe$^{+}$, Fe$^{2+}$, and Fe$^{3+}$ ions using photon-ion merged-beams measurements at a synchrotron light source and accompanying theoretical calculations. Results for Fe$^+$ and Fe$^{3+}$ ions have already been published by \citet{Schippers2017} and \citet{Beerwerth2019}, respectively. The photoabsorption cross-section of neutral Fe$^0$ had already been measured previously by  \citet{Richter2004}.

In this work, we present our measurements of relative cross sections for up to five fold ionization of Fe$^{2+}$ ions via ionization or excitation of the $L$-shell. These data provide, in particular, accurate information on the positions and shapes of photoionization resonances associated with the excitation of a $2p$ electron. The data will also facilitate a reliable identification of Fe$^{2+}$ absorption features in, for example,  astrophysical x-ray spectra. Furthermore,  extensive quantum theoretical calculations to simulate the experimental spectra and to identify the dominant decay channels were performed. These results are vital for determining the charge balance and elemental abundance in astrophysical plasmas.

The Fe$^+$ and Fe$^{3+}$ papers \citep{Schippers2017,Beerwerth2019} provide an extended motivation for these investigations and a comprehensive discussion of the related literature. Therefore, the present paper confines itself primarily to the aspects that are specific for $L$-shell photoionization and photoabsorption of Fe$^{2+}$. Previous work on $L$-shell photoionization of Fe$^{2+}$ ions reported theoretical calculations of cross sections for direct photoionization of a $L$-shell electron \citep{Reilman1979a,Verner1993a} and theoretical calculations of the deexcitation cascades that evolve after the removal of a $2s$ or a $2p$ electron \citep{Kaastra1993,Kucas2019,Kucas2020}.

\section{Experiment}\label{sec:exp}

The experiment was performed at the end station PIPE \citep{Schippers2014,Mueller2017,Schippers2020} of the photon beamline P04 \citep{Viefhaus2013} at the synchrotron light source PETRA III which is operated by DESY in Hamburg, Germany. As for our previous work on $L$-shell photoionization of Fe$^{+}$ and Fe$^{3+}$ \citep{Schippers2017,Beerwerth2019}, we have employed the photon-ion merged-beams technique  \citep[for recent overviews see ][]{Schippers2016,Schippers2020c} to measure cross sections  for single and multiple photoionization of Fe$^{2+}$ ions. The experimental photon-energy range was 690--920~eV. The photon-energy bandwidth was about 1.0~eV corresponding to the full-width-at-half-maximum (FWHM) of a Gaussian distribution function. The maximum energy-dependent photon flux was $8\times10^{13}$~s$^{-1}$. For an accurate determination of the photon energy scale, the same calibration was used as for the measurements with Fe$^+$ \citep{Schippers2017}, taking into account the differences in the Doppler shift between the faster Fe$^{2+}$ ions and the slower Fe$^+$ ions. The remaining one-sigma uncertainty of the experimental photon-energy scale is $\pm$0.2~eV.

\begin{figure}
\includegraphics[width=\columnwidth]{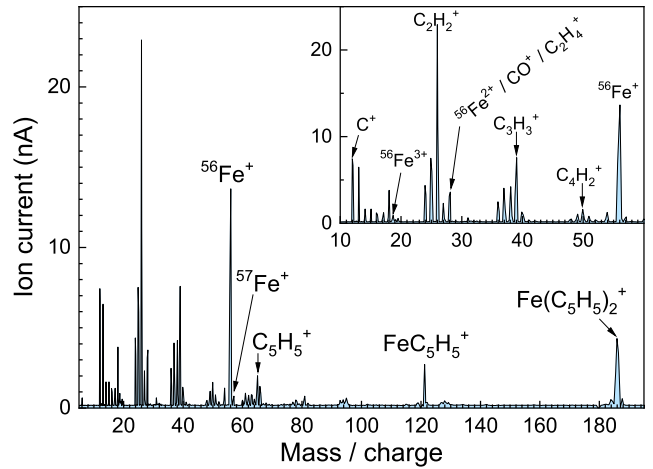}
\caption{\label{fig:mass} Measured primary-ion mass/charge spectrum. The inset enlarges the region of the spectrum that contains the Fe$^{+}$, Fe$^{2+}$, and Fe$^{3+}$ signals.}
\end{figure}

The Fe$^{2+}$ ion beam was produced by leaking ferrocene, Fe(C$_5$H$_{5}$)$_2$, vapor into an electron-cyclotron resonance (ECR) ion source \citep{Schlapp1995a} operated on an electrostatic potential of 6~kV. The extracted ion beam was mass/charge analyzed by passing it through a double-focussing dipole magnet. Figure~\ref{fig:mass} shows the composition of the ion beam as a function of mass-to-charge ratio with mass measured in atomic mass units and charge in the elementary charge unit. This mass/charge spectrum was obtained by scanning the magnetic field strength of the analyzing magnet and simultaneously recording the ion current collected in a Faraday cup. The $^{56}$Fe$^{2+}$ signal occurs at a mass-to-charge ratio of 28. Other species, such as CO$^+$ and C$_2$H$_4^+$ are also expected to contribute to the measured ion current at this mass/charge ratio. Pure Fe$^{2+}$ beams could be obtained for the isotope $^{57}$Fe, but only at a much reduced ion current due to the relatively low natural abundance of this isotope of only 2.1\% \citep{Meija2016}. The $^{56}$Fe$^{2+}$ (including contaminants) and $^{57}$Fe$^{2+}$ ion currents were up to 27~nA and 0.12 nA, respectively.

As described previously for single and multiple ionization of Fe$^+$ ions \citep{Schippers2017}, relative cross sections $\sigma_m$ for $m$-fold ionization ($m=1-5$) were measured individually for each product ion Fe$^{q+}$ with charge state $q=m+2$. To this end, the product ions were magnetically separated from the primary ion beam and directed onto a detector operated in the single-particle counting mode. This charge separation ensured that any photoionized products from the  contaminants of the $^{56}$Fe$^{2+}$ primary ion beam did not reach the product-ion detector.  For most of the present measurements, the potentially contaminated $^{56}$Fe$^{2+}$ beam was used, as it delivered much higher photoionization signals than those obtained using a pure $^{57}$Fe$^{2+}$ beam. Relative cross sections were measured with the isotope $^{57}$Fe at a few selected photon energies and the $^{56}$Fe$^{2+}$ data were scaled to match the $^{57}$Fe$^{2+}$ relative cross sections.

\begin{figure*}
\includegraphics[width=\textwidth]{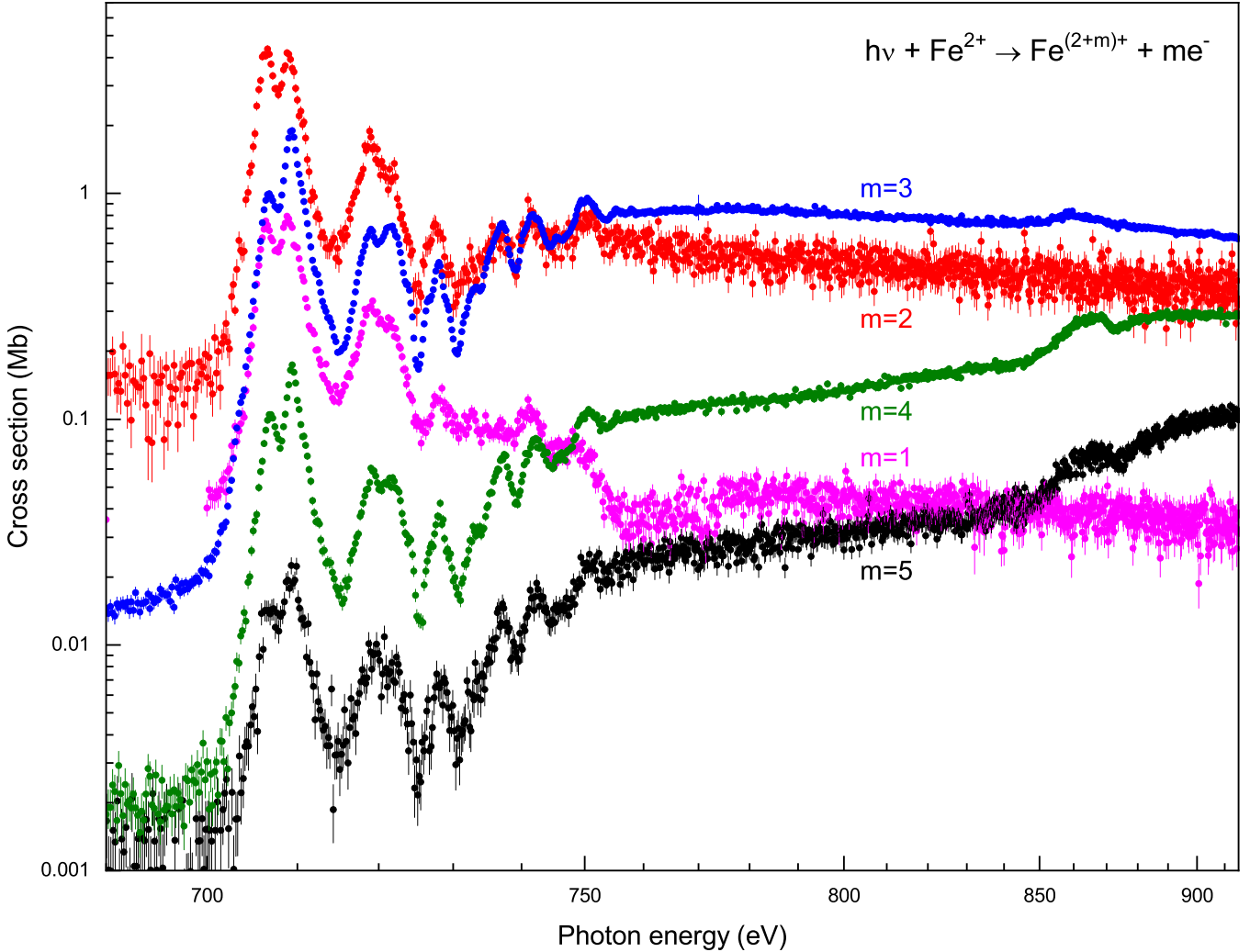}
\caption{\label{fig:all}Measured partial cross sections $\sigma_m$ for m-fold photoionization of Fe$^{2+}$. For a better view of the low-energy resonance structures, the energy scale is compressed towards high photon energies according to the formula $E' = \sqrt{\log \left(E - 600 \, \mathrm{~eV} \right)}$. The absolute cross-section scale (1~Mb = $10^{-18}$~cm$^2$) was obtained by scaling the cross section sum (Equation~\ref{eq:sum}) to the theoretical cross section for photoabsorption (see text).}
\end{figure*}

In principle, the PIPE setup permits photoionization cross sections to be placed on an absolute scale, as was achieved in our work on photoionization of Fe$^+$ ions \citep{Schippers2017}. This requires measuring the spatial profiles of the ion beam and the photon beam, from which the geometrical beam overlap factor can be obtained. For Fe$^{3+}$, such measurements could not be carried out because of a technical problem that could not be solved within the allocated beamtime \citep{Beerwerth2019}. Since the Fe$^{2+}$ and the Fe$^{3+}$ data were taken during the same beamtime, absolute cross sections could not be measured for Fe$^{2+}$ as well. Following the same approach as for the Fe$^{3+}$ data, we multiplied the relative cross sections by a common factor such that the cross-section sum

\begin{equation}\label{eq:sum}
 \sigma_\Sigma = \sum_{m = 1}^{5} \sigma_m
\end{equation}

\noindent matches the theoretical absorption cross section of \citet{Verner1993a} at 690~eV (see below). The implicit assumption that $\sigma_\Sigma$ in Equation~\ref{eq:sum} represents the Fe$^{2+}$ absorption cross section is justified because all significant reaction channels were measured.  This normalization procedure was motivated by the fact that the absolute experimental cross section for photoabsorption of Fe$^+$  agrees with the corresponding theoretical cross section of \citet{Verner1993a} within the $\pm 15\%$ total experimental uncertainty at a 90\% confidence level \citep{Schippers2017}.

\section{Theory} \label{sec:theo}

For a deeper insight into the experimental findings, we performed theoretical calculations using the relativistic Multi-Configuration Dirac-Fock (MCDF) method \citep{Grant2007} and the Hartree-Fock method with relativistic extensions  \citep[HFR,][]{Cowan1981}. The use of these methods has been described more extensively in our publications on photoionization of Fe$^+$ and Fe$^{3+}$ \citep{Schippers2017,Beerwerth2019}. Here, these methods were used to calculate Fe$^{2+}$ absorption cross sections, accounting for direct ionization of a $2p$ electron and for the excitation of a $2p$ electron from the Fe$^{2+}$($1s^2\,2s^2\,2p^6\,3s^2\,3p^6\,3d^6$) ground configurations to higher $nd$ and $n's$ subshells. We included $3\leq n \leq 5$ and $4\leq n' \leq 5$ in the MCDF calculations. In the HFR calculations, we have only considered $2p\to3d$ excitations and  direct $2p$ ionization, the latter of which was not included in the MCDF calculations.

The MCDF computational tools allow for the  modelling of the deexcitation pathways due to Auger cascade processes that accompany the initial creation of a $2p$ hole. Since the focus of our investigations is on the resonance features that are associated with $2p$ excitation, we did not consider cascades initiated by $2s$-hole creation. We used the \textsc{Grasp2k} \citep{Joensson2007a} program package to generate the approximate wave functions and employed the tools of the \textsc{Ratip} code \citep{Fritzsche2001, Fritzsche2012a} to compute the required cross sections and transition rates. The same approach was used for Fe$^{2+}$ as was used in our work on Fe$^{3+}$ \citep{Beerwerth2019}, where more details are given. Briefly, the cascade  includes all energetically allowed (two-electron) Auger processes where one of the electrons fills a lower subshell and another is ejected from the ion.  In addition we accounted for three-electron Auger decays where a third electron undergoes a shake-down transition. The inclusion of these generally weak decay channels has been found to be essential to explain the highest product charge states obtained experimentally \citep{Schippers2016,Schippers2017,Beerwerth2019}. In order to keep the computations manageable, we assume that the radiative losses are negligible, i.e., that all levels that are energetically allowed to autoionize will do so. Cascades initiated by  direct $L$-shell ionization of Fe$^{2+}$ were previously calculated by \citet{Kaastra1993} and more recently by \citet{Kucas2019,Kucas2020}. In Section~\ref{sec:cascades} below, we compare the results from these earlier studies with the present calculations.

\begin{figure}
\includegraphics[width=\columnwidth]{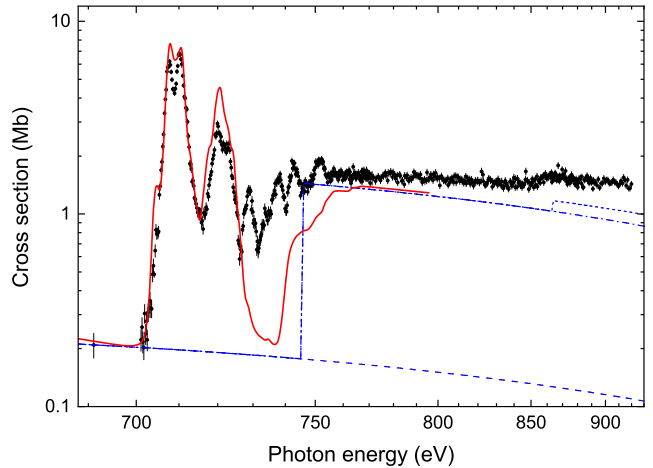}
\caption{\label{fig:abs} Experimental total absorption cross-section $\sigma_\Sigma$ (Equation~\ref{eq:sum}, symbols) and the theoretical subshell-resolved cross sections $\sigma_{nl}$ for photoionization by \citet[dashed line: $\sigma_{3d}+\sigma_{3p}+\sigma_{3s}$, dash-dotted line: $\sigma_{3d}+\sigma_{3p}+\sigma_{3s}+\sigma_{2p}$, short-dashed line: $\sigma_{3d}+\sigma_{3p}+\sigma_{3s}+\sigma_{2p}+\sigma_{2s}$]{Verner1993a} with the calculated thresholds for direct $2p$ and $2s$ ionization occurring at 745.1 and 863.1~eV, respectively. The full line represents the result of the present HFR calculation assuming an initial statistical population of the various levels of the $3d^6$ ground configuration and convoluted with a Gaussian of 1.0~eV FWHM. As in Figure~\ref{fig:all}, the energy scale is compressed for large energies to enhance the visibility of the low-energy resonance structures.}
\end{figure}

\section{Results and discussion} \label{sec:res}

\begin{deluxetable*}{llllllll}
\tablecaption{\label{tab:Xsec} Measured cross sections $\sigma_m$  for $m$-fold photoionization of Fe$^{2+}$ ions (Figure~\ref{fig:all}), resulting sum cross section $\sigma_\Sigma$ (Equation~\ref{eq:sum}, Figure~\ref{fig:abs}), and mean product charge state $\overline{q}$ (Equation~\ref{eq:qmean}, Figure~\ref{fig:frac}b). The numbers in parentheses provide the one-sigma statistical experimental uncertainties. The systematic uncertainties of the photon-energy scale is $\pm0.2$~eV. The cross sections were put on an absolute scale by scaling $\sigma_\Sigma$ to the theoretical absorption cross section of \cite{Verner1993a} below 700~eV (see text for details).}
\tablehead{
    \colhead{Energy~(eV)} &
    \colhead{$\sigma_{1}$~(Mb)\tablenotemark{a}}&
    \colhead{$\sigma_{2}$~(Mb)}&
    \colhead{$\sigma_{3}$~(Mb)}&
    \colhead{$\sigma_{4}$~(Mb)}&
    \colhead{$\sigma_{5}$~(Mb)}&
    \colhead{$\sigma_{\Sigma}$~(Mb)} &
    \colhead{$\bar q$}
    }
\startdata
691.062	& 0.0359(40) & 0.155(30)   &	0.01454(11) &	0.00178(36) &	0.00203(59) &	0.209(31) & 3.94(73) \\
701.077	& 0.0404(33) &	0.153(28) &	0.0238(12) &	0.00303(47) &	0.00202(58) &	0.222(28) & 3.98(63)  \\
710.091	& 0.771(15)  &	4.15(15)  & 	1.7580(80) &	0.1421(32) &	0.0158(16) & 	6.84(16) &	4.19(11) \\
720.107	& 0.315(10)  & 1.773(97)   & 	0.6945(56) &	0.0607(20) &	0.0085(12) &	2.852(97) & 	4.18(17)  \\
750.153	& 0.0760(46) &	0.801(66) &	0.8789(66) & 	0.0959(25) & 	0.0214(18) & 	1.873(67) &	4.56(18) \\
799.829	& 0.0416(47) & 	0.530(59) &	0.820(20)  & 	0.1291(28) & 	0.0325(27) & 	1.554(62) & 	4.73(21) \\
849.906	& 0.0376(47) & 	0.402(51) & 0.735(18)  & 	0.1936(34) & 	0.04688(33) &	1.415(55) & 4.87(20) \\
899.783	& 0.0344(47) & 	0.388(51) & 0.6716(56) &	0.2852(68) & 	0.0904(47)	& 	1.469(52) &		5.01(17) \\
\enddata
\tablenotetext{a}{1~Mb = $10^{-18}$~cm$^2$}
\tablenotetext{}{(This table is available in its entirety in machine-readable form.)}
\end{deluxetable*}

Our measured cross sections $\sigma_m$ ($1 \leq m \leq 5$) for single and multiple (up to five-fold) photoionization of Fe$^{2+}$ ions  are plotted in Figure~\ref{fig:all} and also provided in Tab.~\ref{tab:Xsec} in numerical form. At photon energies of 690--750~eV, the dominant process is double ionization ($m=2$).  Up to $\sim$749~eV, a bound-bound transition forms a hole that can relax by emission of one or more electrons (with two being the dominant mode).  Above $\sim$749~eV, a bound-free transition changes the charge state by one and forms a hole that can relax by emission of one or more electrons (with two being the dominant mode). In the theoretical cross sections of \citet{Verner1993a}, this threshold for direct $2p$ ionization occurs at 745.1~eV (Figure~\ref{fig:abs}). The most obvious signature of this threshold in our data is the step-like decrease of the single-ionization cross section (magenta symbols in Fig.~\ref{fig:all}) that sets in for energies $\gtrsim$749~eV. Below this threshold, a $2p$ electron can only be resonantly excited to a higher partially occupied or unoccupied subshell, i.e, the primary photon-ion interaction does not change the charge state of the ion. At higher energies above the $2p$ threshold, a $2p$ electron can be directly ionized. This primary process increases the ion charge state by one. In both cases, a $2p$ hole is created. The subsequent deexcitation of the multiply excited $2p$-hole configurations proceeds via a cascade of autoionization and radiative processes that produces the observed distributions over the various measured product charge states as already sketched in Section~\ref{sec:theo} and as will be discussed in more detail below.

The experimental absorption cross section (Equation~\ref{eq:sum}) is displayed in Figure~3, together with the theoretical absorption cross section of \citet{Verner1993a}. As explained above, the experimental cross-section scale was normalized to the theoretical one at a photon energy of about 690~eV. The theoretical cross section of \citet{Verner1993a} does not contain any contribution from resonant photoionization. Thus, it clearly displays the calculated $2p$ and $2s$ thresholds at 745.1 and 863.1~eV, respectively. Comparatively weak signatures for the $2s$ ionization threshold can also be seen in the experimental data in Figs.~\ref{fig:all} and \ref{fig:abs}. However, in the experimental cross sections the thresholds are blurred by resonances associated with $2p$ and $2s$ excitation.

The dominant features in all measured cross sections are the two broad resonance structures at around 708 and 720 eV which can be attributed to $2p\to3d$ excitations. The $\sim$12~eV energy difference between these two structures corresponds to the $2p_{1/2,\: 3/2}$ spin-orbit splitting. This splitting is reproduced by our present HFR calculations which are also displayed in Figure~\ref{fig:abs}. As already mentioned, the HFR calculations account for $2p\to 3d$ excitations, but do not include the weaker resonances at higher energies, which are predominantly associated  with $2p\to nd$ excitations to higher subshells with $n \geq 4$. Hence, the HFR results lie below the measurements from $\sim$730--760~eV. The HFR calculations also account for direct $2p$ ionization, which is the dominant contribution to the Fe$^{2+}$ absorption cross section at energies above 760 eV and continuing  up to the highest experimental photon energy studied here.

\begin{figure}
\centering
\includegraphics[width=1.0\linewidth]{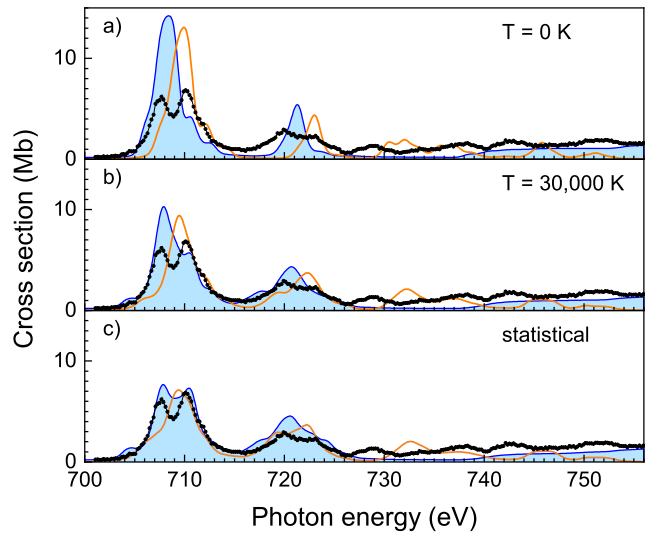}
\caption{Computed absorption cross sections for different populations of the $3d^6$ ground configuration compared with the experimental  cross-section sum (full symbols) from Figure~\ref{fig:abs}. The blue shaded and the orange full curves are the results of our HFR and MCDF calculations, which have been shifted by -2.5 and -1.0~eV, respectively,  in order to align (by eye) the computed resonance structures with the experimental results. All theoretical cross section were convolved with a Gaussian with a FWHM of 1.0 eV in order to account for the experimental photon-energy bandwidth.}
\label{fig:thcomp}
\end{figure}

The Fe$^{2+}$([Ar]\,$3d^6$) ground configuration splits into 34 fine-structure levels which span an energy range of $\sim$7~eV \citep{Kramida2019}. It must be assumed that all excited levels were populated in the ion source. The lifetimes of these levels are sufficiently long that the excited ions partly survive the transport from the ion source to the photon-ion interaction region. Consequently, the ion beam consisted of a mixture of ions in different levels.  Our MCDF and HFR calculations were carried out for all individual levels of the ground configuration. As shown in Figure~\ref{fig:thcomp}, different assumptions on the initial level population in the ion beam lead to significantly different results for the calculated absorption cross sections. In Figure~\ref{fig:thcomp}a we assumed a pure ground-level population. Figure~\ref{fig:thcomp}b shows calculations for a Boltzmann distribution of energy levels for a temperature of 30,000~K.  Figure~\ref{fig:thcomp}c contains theoretical results for a statistical mixture of all levels pertaining to the Fe$^{2+}$ ground configuration, where where the population of each level is weighted by its degeneracy. Apparently, the latter provides the best agreement with the experimental absorption cross section, if the theoretical resonance positions are uniformly shifted by -2.5 and -1.0~eV in case of the HFR and the MCDF calculations, respectively. Similar shifts were applied to the corresponding theoretical results for Fe$^+$ and Fe$^{3+}$ \citep{Schippers2017,Beerwerth2019}.

As already mentioned, the HFR calculations account for direct $2p$ ionization and $2p\to3d$ excitation. Under the assumption of a statistical initial level distribution, the calculated cross section agrees rather well with the experimental absorption cross section in the energy range of the photoionization resonances that are associated with the $2p\to3d$ excitation. The resonances above $\sim$726~eV are related to the excitation of a $2p$ electron to higher subshells such as the $2p\to4d$ and $2p\to5d$ excitations that were included in the MCDF calculations. But for the MCDF calculations, the agreement between the computed and experimental resonance structure is less satisfying, as compared to the HFR calculations (Figure~\ref{fig:thcomp}). This is attributed to the rather limited consideration of configuration interaction by the present MCDF calculations, which were more geared towards the simulation of the deexcitation cascades that set in after the initial creation of the $2p$ hole and that are discussed next.

\subsection{Cascade Calculations} \label{sec:cascades}

From the measured cross sections $\sigma_m$ for $m$-fold photoionization, the product charge-state fractions, i.e., the probabilities of an atom to end up in charge state $q$, can be derived as

\begin{equation}\label{eq:fq}
f_q\left(E_{\mathrm{ph}}\right) = \frac{\sigma_q}{\sigma_\Sigma},
\end{equation}

\noindent where $\sigma_\Sigma$ is given by Equation~\ref{eq:sum}. A key feature of the quantities $f_q$ is, that the systematic uncertainty of the absolute cross section scale cancels out. The fractions $f_q$ can also be used for the calculation of the mean product-charge state

\begin{equation}\label{eq:qmean}
\bar q\left( E_{\mathrm{ph}} \right) = \sum_{q = 3}^7 q f_q = \frac{1}{\sigma_\Sigma} \sum_{m = 1}^5 \left(m + 2 \right) \sigma_m.
\end{equation}

\begin{figure}[t]
\includegraphics[width=\linewidth]{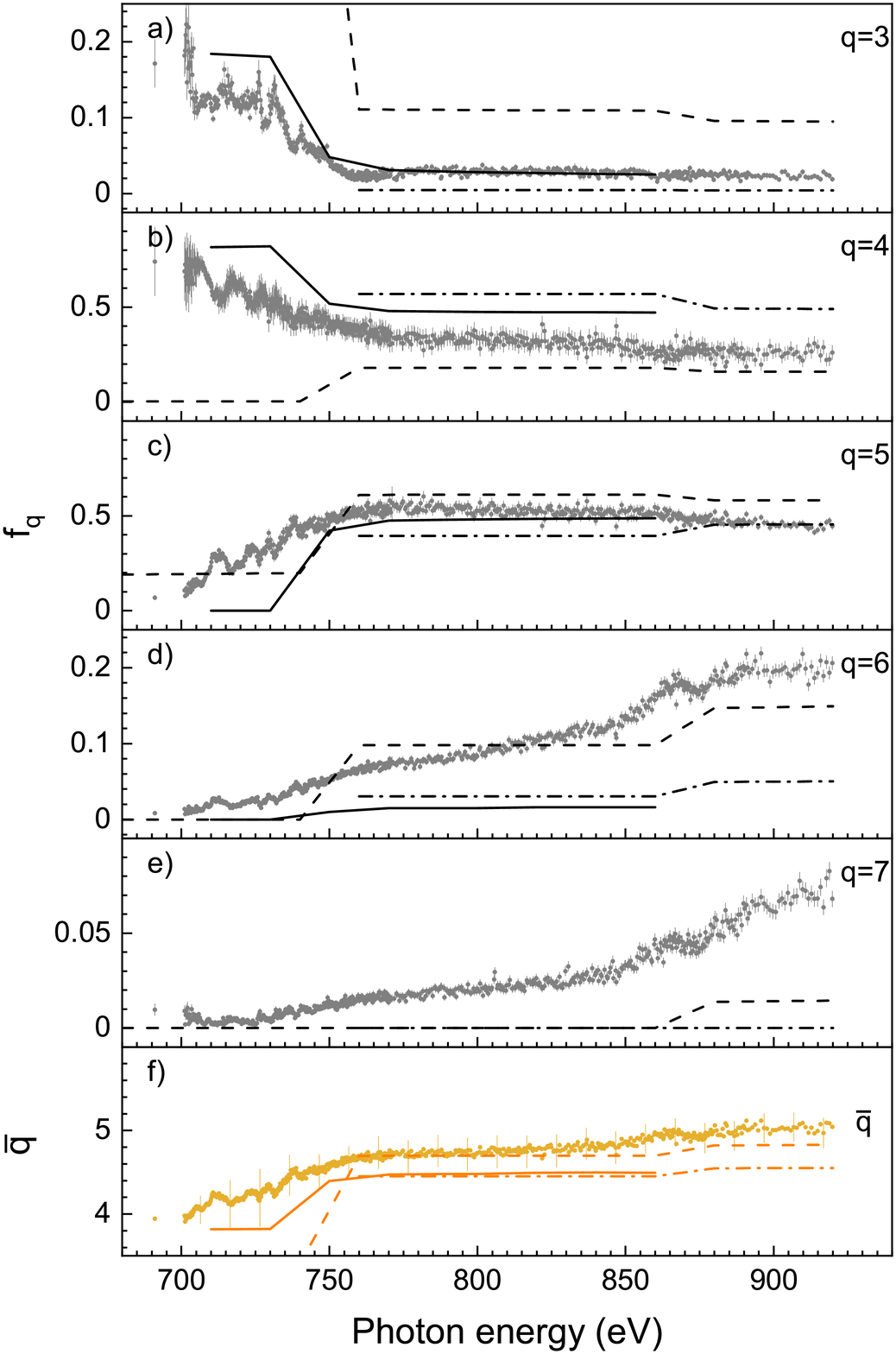}
\caption{\label{fig:frac} Product charge-state fractions $f_q$ populated by photoionization of Fe$^{2+}$ for panel (a) $q=3$, (b) $q=4$, (c) $q=5$, (d) $q=6$, and (e) $q=7$: Experimental results (grey symbols) are compared to our MCDF computations (full lines) and to the theoretical results by \citet[dashed lines]{Kaastra1993} and by \citet[dash-dotted lines]{Kucas2019,Kucas2020}, both weighted by the subshell-specific photoionization cross sections of \citet{Verner1993a} that are displayed in Figure~\ref{fig:abs}. There is no full line in panel (e) since our MCDF cascade model does not predict any seven-fold charged product ions. Panel (f): Mean charge state derived from the experimental data (orange symbols) and our cascade calculations (full line). The dashed lines and dash-dotted lines result from the subshell specific charge fractions of \citet{Kaastra1993} and of \citet{Kucas2019,Kucas2020}, respectively, in combination with the subshell specific ionization cross sections of \citet{Verner1993a}.}
\end{figure}

Figure~\ref{fig:frac} shows our experimental and theoretical product charge-state fractions following Fe$^{2+}$ photoionization. Our MCDF calculations only account for cascades that follow the initial creation of a $2p$, $3s$, or $3p$ hole. Therefore, our theoretical values only apply to energies below the threshold for direct $2s$ ionization at about 860~eV. The calculated product charge-state fractions follow the experimental values reasonably well considering the simplifications that were applied in order to keep the computations tractable. This is particularly true for the lower product charge states $q=3$ and $4$. The theoretical values for $q=5$ agree with the corresponding experimental quantities only for energies above the threshold for direct $2p$ ionization at about 750~eV. At lower energies the MCDF values for $q=5$ are much too small (they are essentially zero).  The MCDF calculations significantly underestimate the production of the product charge state $q=6$ over the entire experimental energy range. The calculations do not predict any sizeable $q=7$ fraction. We attribute this increasing failure of our MCDF cascade model with increasing charge state to a mismatch of the computed (auto-)ionization thresholds and to the neglect of many-body processes in the cascade calculations beyond the three-electron processes mentioned in Section~\ref{sec:theo}. At present, such detailed calculations cannot be easily carried out for complex ions such as Fe$^{2+}$ within our MCDF framework, although more advanced cascade computations, based on detailed cascade trees, are under way \citep{Fritzsche2019}.

Earlier work on cascades, after inner-shell ionization of Fe$^{2+}$, has been carried out by \citet{Kaastra1993} and more recently by \citet{Kucas2019,Kucas2020}. These authors calculated the product charge-state distributions resulting from the removal of an inner-shell electron and the subsequent deexcitation cascades.  In order to compare these results with the measured product charge-state fractions, the individual distributions for each inner-shell hole have to be weighted by the corresponding cross sections for inner-shell ionization \citep{Schippers2017,Beerwerth2019}. To this end, we used the subshell specific photoionization cross sections of \citet{Verner1993a} that are displayed in Figure~\ref{fig:abs}. The resulting product charge-state fractions are also displayed in Figure~\ref{fig:frac}. The overall agreement of the results of  \citet{Kaastra1993} with the experimental findings is not as good as compared to our MCDF calculations, in particular for the lowest two product-ion charge states. For the higher product charge states the agreement is  somewhat better. \citet{Kucas2019,Kucas2020} have only considered the production of initial $2p$ and $2s$ holes. Therefore, their values  can only be applied at energies above the $2p$ ionization threshold and are biased towards higher product charge states. They significantly underestimate the production of Fe$^{6+}$. The \citet{Kucas2019,Kucas2020} results for $q=4$ and $5$ are in reasonable agreement with our MCDF and experimental results, given the various limitations of the theories.  The agreement with the \citet{Kaastra1993}  results is only reasonable for $q=5$. Single ionization is severely underestimated by \citet{Kucas2019,Kucas2020}.

None of the above discussed cascade calculations reproduces the experimental findings for the product charge-state fractions $f_q$ in all the details. This is likely due to the considerable complexity of the problem under consideration which, currently, can only be treated by making simplifying assumptions and by neglecting higher order processes such as direct double ionization which probably form a wide continuum of cross-section contributions. Nevertheless, as shown in Figure~\ref{fig:frac}, all methods agree reasonably well with the experimental findings (Tab.~\ref{tab:Xsec}) for the photon-energy dependent mean charge state $\bar{q}$ (Equation~\ref{eq:qmean}).

\subsection{Summary of the experimental and theoretical absorption cross-sections  for neutral and low-charged iron}

\begin{figure}[t]
\includegraphics[width=\linewidth]{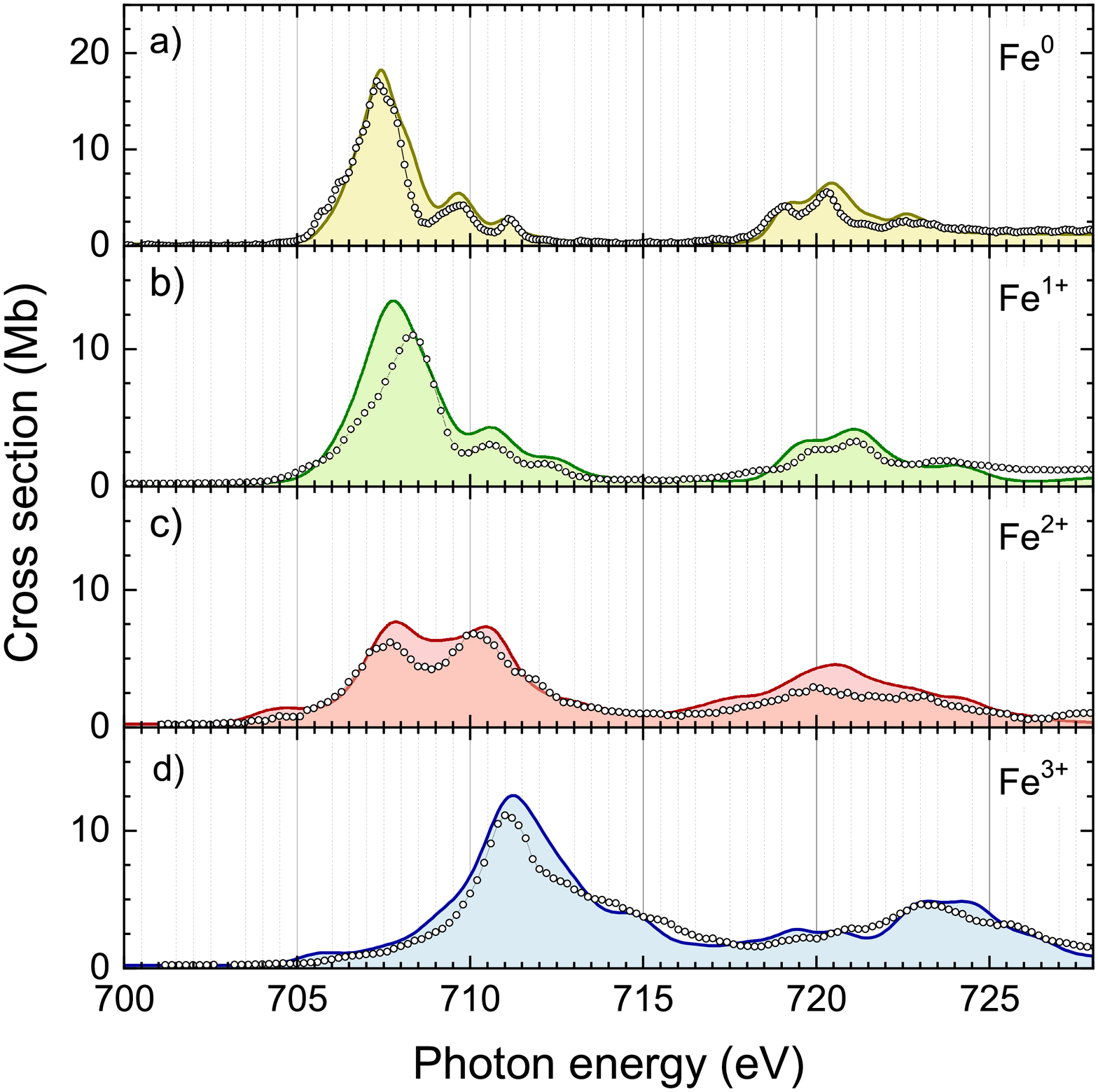}
\caption{\label{fig:Fe0123}Experimental (symbols) and HFR (full shaded curves) photoabsorption cross sections of a) Fe$^0$, b) Fe$^{+}$, c) Fe$^{2+}$, and d) Fe$^{3+}$. The Fe$^0$, Fe$^+$ and Fe$^{3+}$ data have been taken from \citet{Richter2004}, \citet{Schippers2017} and \citet{Beerwerth2019}, respectively. The theoretical cross sections are for statistical mixtures of ground-configuration levels and account for the experimental photon-energy bandwidth which was $\sim$0.6~eV for Fe$^0$ and  $\sim$1~eV for Fe$^+$, Fe$^{2+}$, and  Fe$^{3+}$. The one-sigma uncertainty of the experimental photon-energy scale is $\pm$0.2~eV for  Fe$^+$, Fe$^{2+}$, and  Fe$^{3+}$ and unspecified for Fe$^0$.}
\end{figure}

The present paper is the last in a sequence of publications on photoionization of low-charged iron ions. Therefore, we can now provide a comparison with our previous work on Fe$^+$ \citep{Schippers2017} and on Fe$^{3+}$ \citep{Beerwerth2019} with a focus on the most prominent features in the iron $L$-shell photoabsorption. In particular, these features should be useful for the identification of gas-phase iron in the ISM. Figure~\ref{fig:Fe0123} displays our experimental and theoretical results for Fe$^+$, Fe$^{2+}$, and Fe$^{3+}$ ions and also the corresponding results for neutral iron atoms of \citet{Richter2004}. In Figure~\ref{fig:Fe0123}, their Fe$^0$ relative experimental cross sections were multiplied by an energy-independent factor to scale them to the theoretical cross-section scale. All the theoretical cross sections that are displayed in this figure are HFR results (see Section~\ref{sec:theo}) for statistical mixtures of ground-configuration levels as discussed in Section~\ref{sec:res} for Fe$^{2+}$. As already mentioned, the theoretical energy scales were shifted by eye in order to best line up the theoretical resonance positions with the experimental ones.  The remaining minor discrepancies between theory and experiment are most probably due to the inherent limitations of the theoretical method and uncertainties in the level populations for the experimental results. In the following we thus use the experimentally benchmarked theoretical cross sections for the comparison with absorption cross sections for solid materials that are expected to be present in interstellar dust particles.

\section{Comparison with other forms of iron}\label{sec:comp}

The nature of the chemical binding of iron to interstellar dust particles is currently being debated. For example, from experiments in microgravity \citet{Kimura2017} concluded that the probability of the formation of pure iron grains should be very low and that, consequently, iron should be bound in iron compounds or accreted as impurities on other grains in the ISM.  Another hypothesis was pursued by \citet{Bilalbegovic2016} who calculated infrared absorption spectra of hydrogenated iron particles.  From a variety of astronomical observations and related astrophysical modeling, several groups  inferred that a large fraction of the iron in the ISM could be incorporated as inclusions in silicate grains \citep{Zhukovska2018,Westphal2019,Zafar2019}. Using an ion implantation technique, \citet{Leveneur2011} produced Fe nanoparticles in silica, which could serve as a proxy for iron locked in interstellar dust particles. In addition, \citet{Lee2009} have carefully measured absorption data for a number of iron compounds.

\begin{figure}[t]
\includegraphics[width=\linewidth]{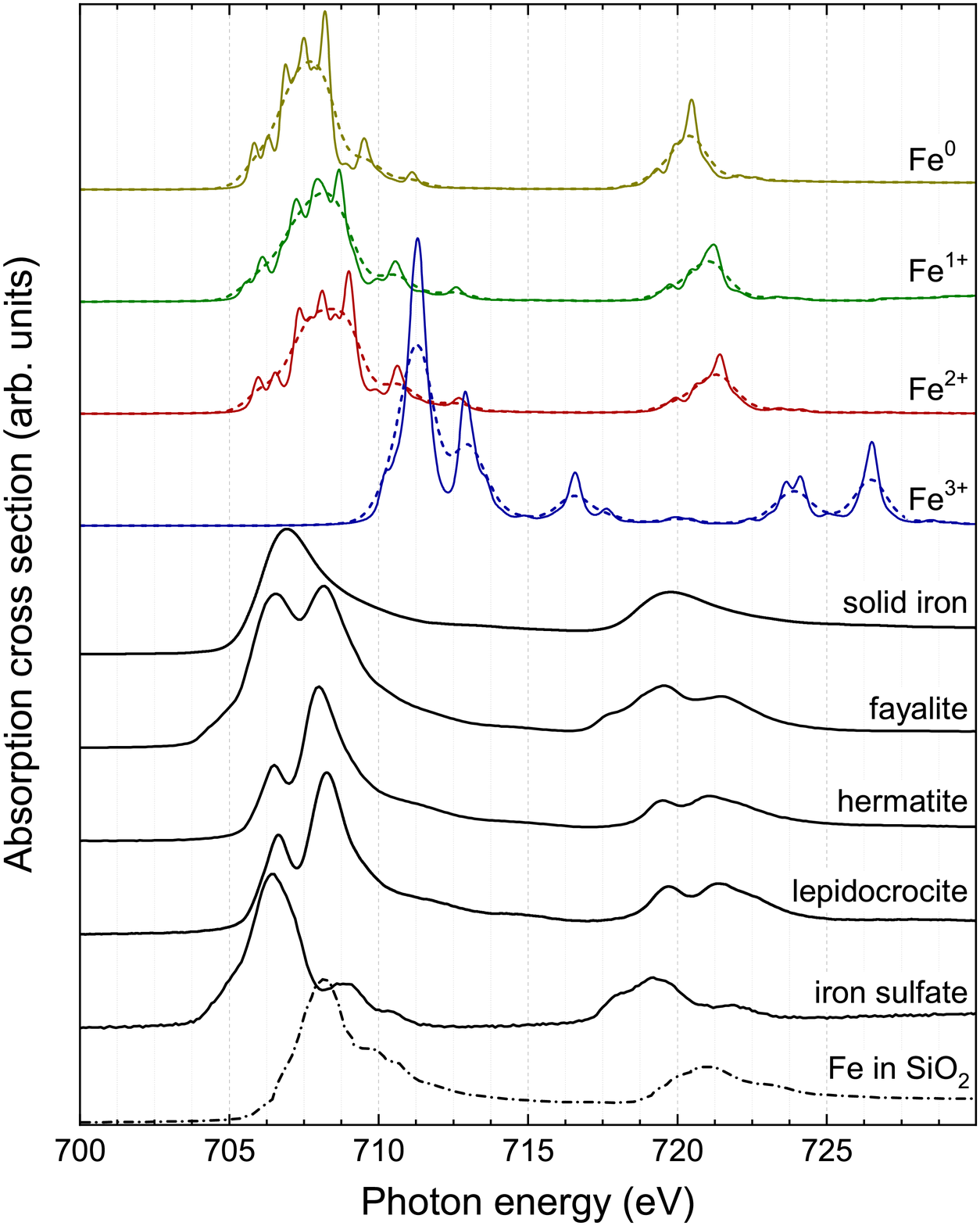}
\caption{\label{fig:Fesolid}Comparison of  photoabsorption cross sections  for gas-phase neutral and low-charged ions with photoabsorption cross sections for solid iron and solid iron-bearing compounds. The cross sections for Fe$^0$, Fe$^+$, Fe$^{2+}$, and Fe$^{3+}$ (colored lines) are the experimentally benchmarked HFR results for photoabsorption of the ground level only. The full and the dashed colored lines were calculated taking into account photon energy spreads of 0.24~eV and 1.0~eV, respectively. The data for solid iron  and iron-bearing compounds have been taken from the work of \citet[][black full lines]{Lee2009} and \citet[][black dash-dotted line]{Leveneur2011}. The one-sigma uncertainty of the experimental photon-energy scale is $\pm$0.2~eV for  Fe$^+$, Fe$^{2+}$, and  Fe$^{3+}$ and unspecified for the other samples (see text).}
\end{figure}

Figure \ref{fig:Fesolid} presents a comparison of the experimentally benchmarked theoretical gas-phase data for neutral and low-charged ions with the absorption cross sections for solid iron and solid iron compounds of  \citet{Lee2009} and \citet{Leveneur2011}. The theoretical gas-phase data are different from those displayed in Figure~\ref{fig:Fe0123} where  statistical populations of several levels were considered in order to account for the high temperatures in the ion source. The temperatures in the ISM are considerably lower such that usually only the ground levels are significantly populated. This is accounted for in the theoretical data for Fe$^0$, Fe$^+$, Fe$^{2+}$, and Fe$^{3+}$ displayed in Figure~\ref{fig:Fesolid}. In addition, a finite photon energy bandpass was considered by convoluting the theoretical cross sections with Gaussians with FWHM of 1.0~eV and 0.24~eV. The latter energy spread corresponds to a resolving power $E/\Delta E = 3000$ as was used in the experiments of \citet{Lee2009}. This resolving power approximately corresponds to what is currently foreseen for the future Athena x-ray telescope \citep{Barret2020}.

When comparing x-ray absorption data from different sources, the calibration uncertainties of the different photon energy scales are an issue of concern. For the Fe$^+$, Fe$^{2+}$, and Fe$^{3+}$ absorption cross sections this uncertainty is $\pm$0.2~eV. In our experiments, absorption features in gases were used as reference standards as discussed in considerable detail by \citet{Mueller2017,Mueller2018c}. Neither \citet{Richter2004} nor \citet{Leveneur2011} provide any information on the calibration uncertainties of their energy scales. \cite{Lee2009} refer their data to theoretically calculated values for the iron $2p$ absorption edges as provided by the computer code of \citet{Brennan1992} which uses theoretical absorption data from \citet{Cromer1970,Cromer1981}. This calibration is problematic for two reasons. First, the experimental data do not exhibit a clear step-like threshold absorption feature since this is masked by the near-threshold photoabsorption resonances. Second, the uncertainties of the theoretical calculations are unknown and may be rather large. For example, in their comprehensive list of x-ray calibration features, \citet{Deslattes2003} quote a difference between theoretical and experimental values for the iron $L_3$ edge of 2~eV.  This value is in line with the $-$2.2 eV energy shift that we applied in benchmarking the present theoretical HFR results for photoabsorption of Fe$^{2+}$. This uncertainty of the solid-state data hampers a quantitative comparison between gas-phase and solid-state absorption data. In any case, the differences will become more apparent at higher spectral resolving powers such as those envisaged for future x-ray telescopes.

\section{Summary and Conclusions}\label{sec:summary}

This publication concludes a sequence of papers  \citep{Schippers2017,Beerwerth2019} on the photoionization of low-charged Fe$^{q+}$ ions with $q$ = 1,2, and 3 that was designed to provide accurate data for resonant $L$-shell photoabsorption in order to enable the identification of these ion species in the ISM medium and their discrimination against solid-phase iron locked up in dust grains. The experimental cross sections for $m$-fold ionization ($m$=1--5) for Fe$^{2+}$ were obtained by using the photon-ion merged-beams technique at a synchrotron light source. The cross-section data exhibit strong absorption features associated with the excitation of a $2p$ electron to higher atomic subshells. The positions and relative peak intensities of these features can serve as clear fingerprints for the various iron charge states investigated. We have also carried out quantum theoretical calculations of photoabsorption by Fe$^{2+}$ ions. Furthermore, we have calculated the deexcitation cascades that follow the creation of a $2p$ hole and that determine the final charge state distributions.

Using our combined experimental and theoretical results, we can generate the photoabsorption data needed to model X-ray observations of the ISM.  The experimental results are particularly important for providing experimental benchmarks for the theoretical energy scale. Because of the many-body nature of the problem, the calculated energies of the $L$-shell absorption features have rather large uncertainties. Thus, our experimental benchmarking is vital for achieving the level of accuracy required for astronomical observations. The energy-shifted theoretical results can then be used to generate the needed photoabsorption cross-section data for cold ions, similar to those expected in the ISM.  The experimental results cannot be directly used due to the initial level distribution of the ions corresponding to a much higher temperature than the cold ISM. A direct comparison of the present gas-phase absorption cross section with absorption data for iron-bearing condensed matter compounds from the literature is also hampered by the unknown uncertainties of the photon-energies in these studies. Future work on solid-state absorption data is needed to address this issue.

Our calculations of the Auger cascades that evolve after the primary creation of a $2p$ hole by excitation or direct ionization yield charge-state fractions that are in reasonable agreement with the experimental findings for the lowest two product-ion charge states, Fe$^{3+}$ and Fe$^{4+}$. This is a clear improvement as compared to earlier work by \citet{Kaastra1993} and by \citet{Kucas2019,Kucas2020}. The higher product-ion charge states are not as well described. This is attributed to the simplifications that had to be made in order to keep the computations tractable. For example, our calculations neglected double shake processes that were found to be  important for an accurate description of cascades in a lighter ion \citep{Schippers2016a}. A further development of the atomic structure codes --- such as currently pursued by \citet{Fritzsche2019} --- is required to properly account for the many-electron contributions and a sufficiently large set of decay paths also for heavier ions.

\begin{acknowledgments}

We thank L\'ia Corrales, Eric Gotthelf, Julia Lee, Tim Kallman, and Frits Paerels for stimulating discussions. We acknowledge DESY (Hamburg, Germany), a member of the Helmholtz Association HGF, for the provision of experimental facilities. Parts of this research were carried out at PETRA\,III and we would like to thank Gregor Hartmann, Frank Scholz, and Jörn Seltmann for assistance in using beamline P04.

This research has been funded in part by the German Federal Ministry for Education and Research (BMBF) within the \lq\lq{}Verbundforschung\rq\rq\ funding scheme under contracts 05K16GUC, 05K16RG1, and 05K16SJA.
S.B. and K.S. acknowledge funding from the Initiative and Networking Fund of the Helmholtz Association through the Young Investigators Group program and the SFB 755, "Nanoscale photonic imaging, project B03".
D.W.S. was supported, in part, by the NASA Astrophysics Research Program and the Astrophysics Data Analysis Program. The participation of RAP in the experiment was supported by funds from the Alexander von Humboldt Foundation.
\end{acknowledgments}


\begin{thebibliography}{}
\expandafter\ifx\csname natexlab\endcsname\relax\def\natexlab#1{#1}\fi
\providecommand{\url}[1]{\href{#1}{#1}}
\providecommand{\dodoi}[1]{doi:~\href{http://doi.org/#1}{\nolinkurl{#1}}}
\providecommand{\doeprint}[1]{\href{http://ascl.net/#1}{\nolinkurl{http://ascl.net/#1}}}
\providecommand{\doarXiv}[1]{\href{https://arxiv.org/abs/#1}{\nolinkurl{https://arxiv.org/abs/#1}}}

\bibitem[{Barret {et~al.}(2020)Barret, Decourchelle, Fabian, Guainazzi, Nandra,
  Smith, \& den Herder}]{Barret2020}
Barret, D., Decourchelle, A., Fabian, A., {et~al.} 2020, AN, 341, 224,
  \dodoi{10.1002/asna.202023782}

\bibitem[{Beerwerth {et~al.}(2019)Beerwerth, Buhr, Perry-Sassmannshausen,
  Stock, Bari, Holste, Kilcoyne, Reinwardt, Ricz, Savin, Schubert, Martins,
  Müller, Fritzsche, \& Schippers}]{Beerwerth2019}
Beerwerth, R., Buhr, T., Perry-Sassmannshausen, A., {et~al.} 2019, ApJ, 887,
  189, \dodoi{10.3847/1538-4357/ab5118}

\bibitem[{Bilalbegovi{\'{c}} {et~al.}(2016)Bilalbegovi{\'{c}},
  Maksimovi{\'{c}}, \& Moha{\v{c}}ek-Gro{\v{s}}ev}]{Bilalbegovic2016}
Bilalbegovi{\'{c}}, G., Maksimovi{\'{c}}, A., \& Moha{\v{c}}ek-Gro{\v{s}}ev, V.
  2016, MNRAS, 466, L14, \dodoi{10.1093/mnrasl/slw226}

\bibitem[{Brennan \& Cowan(1992)}]{Brennan1992}
Brennan, S., \& Cowan, P.~L. 1992, RScI, 63, 850, \dodoi{10.1063/1.1142625}

\bibitem[{Cowan(1981)}]{Cowan1981}
Cowan, R.~D. 1981, The theory of atomic structure and spectra (Berkeley:
  California University Press), \dodoi{10.1525/9780520906150}

\bibitem[{Cromer \& Liberman(1970)}]{Cromer1970}
Cromer, D.~T., \& Liberman, D. 1970, JChPh, 53, 1891, \dodoi{10.1063/1.1674266}

\bibitem[{Cromer \& Liberman(1981)}]{Cromer1981}
Cromer, D.~T., \& Liberman, D.~A. 1981, AcCrA, 37, 267,
  \dodoi{10.1107/S0567739481000600}

\bibitem[{Deslattes {et~al.}(2003)Deslattes, Kessler, Indelicato, de~Billy,
  Lindroth, \& Anton}]{Deslattes2003}
Deslattes, R.~D., Kessler, E.~G., Indelicato, J.~P., {et~al.} 2003, RvMP, 75,
  35, \dodoi{10.1103/RevModPhys.75.35}

\bibitem[{Fritzsche(2001)}]{Fritzsche2001}
Fritzsche, S. 2001, JESRP, 114-116, 1155, \dodoi{10.1016/S0368-2048(00)00257-7}

\bibitem[{Fritzsche(2012)}]{Fritzsche2012a}
---. 2012, CoPhC, 183, 1525, \dodoi{10.1016/j.cpc.2012.02.016}

\bibitem[{Fritzsche(2019)}]{Fritzsche2019}
---. 2019, CoPhC, 240, 1, \dodoi{10.1016/j.cpc.2019.01.012}

\bibitem[{Grant(2007)}]{Grant2007}
Grant, I.~P. 2007, Relativistic Quantum Theory of Atoms and Molecules: Theory
  and Computation (New York: Springer), \dodoi{10.1007/978-0-387-35069-1}

\bibitem[{J\"{o}nsson {et~al.}(2007)J\"{o}nsson, He, Froese-Fischer, \&
  Grant}]{Joensson2007a}
J\"{o}nsson, P., He, X., Froese-Fischer, C., \& Grant, I.~P. 2007, CoPhC, 177,
  597, \dodoi{10.1016/j.cpc.2007.06.002}

\bibitem[{Kaastra \& Mewe(1993)}]{Kaastra1993}
Kaastra, J.~S., \& Mewe, R. 1993, A\&AS, 97, 443

\bibitem[{Kimura {et~al.}(2017)Kimura, Tanaka, Nozawa, Takeuchi, \&
  Inatomi}]{Kimura2017}
Kimura, Y., Tanaka, K.~K., Nozawa, T., Takeuchi, S., \& Inatomi, Y. 2017, SciA,
  3, \dodoi{10.1126/sciadv.1601992}

\bibitem[{Kortright \& Kim(2000)}]{Kortright2000}
Kortright, J.~B., \& Kim, S.-K. 2000, PhRvB, 62, 12216,
  \dodoi{10.1103/PhysRevB.62.12216}

\bibitem[{Kramida {et~al.}(2019)Kramida, Ralchenko, Reader, \&
  Team}]{Kramida2019}
Kramida, A., Ralchenko, Y., Reader, J., \& Team, N.~A. 2019, {NIST} atomic
  spectra database (version 5.7.1), [online]. available:
  http://physics.nist.gov/asd, Tech. rep., National Institute of Standards and
  Technology, \dodoi{10.18434/T4W30F}

\bibitem[{Ku{\v{c}}as {et~al.}(2020)Ku{\v{c}}as, Drabu{\v{z}}inskis, \&
  Jonauskas}]{Kucas2020}
Ku{\v{c}}as, S., Drabu{\v{z}}inskis, P., \& Jonauskas, V. 2020, ADNDT, 101357,
  \dodoi{10.1016/j.adt.2020.101357}

\bibitem[{Ku{\v{c}}as {et~al.}(2019)Ku{\v{c}}as, Drabu{\v{z}}inskis,
  Kynien{\.{e}}, Masys, \& Jonauskas}]{Kucas2019}
Ku{\v{c}}as, S., Drabu{\v{z}}inskis, P., Kynien{\.{e}}, A., Masys, {\v{S}}., \&
  Jonauskas, V. 2019, JPhB, 52, 225001, \dodoi{10.1088/1361-6455/ab46fa}

\bibitem[{Lee {et~al.}(2009)Lee, Xiang, Ravel, Kortright, \&
  Flanagan}]{Lee2009}
Lee, J.~C., Xiang, J., Ravel, B., Kortright, J., \& Flanagan, K. 2009, ApJ,
  702, 970, \dodoi{10.1088/0004-637X/702/2/970}

\bibitem[{Leveneur {et~al.}(2011)Leveneur, Waterhouse, Kennedy, Metson, \&
  Mitchell}]{Leveneur2011}
Leveneur, J., Waterhouse, G. I.~N., Kennedy, J., Metson, J.~B., \& Mitchell, D.
  R.~G. 2011, JPCC, 115, 20978, \dodoi{10.1021/jp206357c}

\bibitem[{Meija {et~al.}(2016)Meija, Tyler, Berglund, Willi, {De Bi\'{e}}vre,
  Gr\"{o}ning, Norman, Irrgeher, Robert, Walczyk, \& Prohaska}]{Meija2016}
Meija, J., Tyler, C.~B., Berglund, M., {et~al.} 2016, PApCh, 88, 293,
  \dodoi{10.1515/pac-2015-0503}

\bibitem[{Miedema \& de~Groot(2013)}]{Miedema2013}
Miedema, P.~S., \& de~Groot, F. M.~F. 2013, JESRP, 187, 32,
  \dodoi{10.1016/j.elspec.2013.03.005}

\bibitem[{M\"{u}ller {et~al.}(2017)M\"{u}ller, Bernhardt, Borovik, Buhr,
  Hellhund, Holste, Kilcoyne, Klumpp, Martins, Ricz, Seltmann, Viefhaus, \&
  Schippers}]{Mueller2017}
M\"{u}ller, A., Bernhardt, D., Borovik, Jr., A., {et~al.} 2017, ApJ, 836, 166,
  \dodoi{10.3847/1538-4357/836/2/166}

\bibitem[{M\"{u}ller {et~al.}(2018)M\"{u}ller, Lindroth, Bari, Borovik~Jr.,
  Hillenbrand, Holste, Indelicato, Kilcoyne, Klumpp, Martins, Viefhaus,
  Wilhelm, \& Schippers}]{Mueller2018c}
M\"{u}ller, A., Lindroth, E., Bari, S., {et~al.} 2018, PhRvA, 98, 033416,
  \dodoi{10.1103/PhysRevA.98.033416}

\bibitem[{Reilman \& Manson(1979)}]{Reilman1979a}
Reilman, R.~F., \& Manson, S.~T. 1979, ApJS, 40, 815, \dodoi{10.1086/190605}

\bibitem[{Richter {et~al.}(2004)Richter, Godehusen, Martins, Wolff, \&
  Zimmermann}]{Richter2004}
Richter, T., Godehusen, K., Martins, M., Wolff, T., \& Zimmermann, P. 2004,
  PhRvL, 93, 023002, \dodoi{10.1103/PhysRevLett.93.023002}

\bibitem[{Schippers {et~al.}(2016{\natexlab{a}})Schippers, Kilcoyne, Phaneuf,
  \& M\"{u}ller}]{Schippers2016}
Schippers, S., Kilcoyne, A. L.~D., Phaneuf, R.~A., \& M\"{u}ller, A.
  2016{\natexlab{a}}, ConPh, 57, 215, \dodoi{10.1080/00107514.2015.1109771}

\bibitem[{Schippers \& M\"uller(2020)}]{Schippers2020c}
Schippers, S., \& M\"uller, A. 2020, Atoms, 8, 45, \dodoi{10.3390/atoms8030045}

\bibitem[{Schippers {et~al.}(2014)Schippers, Ricz, Buhr, Borovik, Hellhund,
  Holste, Huber, Sch\"{a}fer, Schury, Klumpp, Mertens, Martins, Flesch, Ulrich,
  R\"{u}hl, Jahnke, Lower, Metz, Schmidt, Sch\"{o}ffler, Williams, Glaser,
  Scholz, Seltmann, Viefhaus, Dorn, Wolf, Ullrich, \&
  M\"{u}ller}]{Schippers2014}
Schippers, S., Ricz, S., Buhr, T., {et~al.} 2014, JPhB, 47, 115602,
  \dodoi{10.1088/0953-4075/47/11/115602}

\bibitem[{Schippers {et~al.}(2016{\natexlab{b}})Schippers, Beerwerth, Abrok,
  Bari, Buhr, Martins, Ricz, Viefhaus, Fritzsche, \&
  M\"{u}ller}]{Schippers2016a}
Schippers, S., Beerwerth, R., Abrok, L., {et~al.} 2016{\natexlab{b}}, PhRvA,
  94, 041401, \dodoi{10.1103/PhysRevA.94.041401}

\bibitem[{Schippers {et~al.}(2017)Schippers, Martins, Beerwerth, Bari, Holste,
  Schubert, Viefhaus, Savin, Fritzsche, \& M\"{u}ller}]{Schippers2017}
Schippers, S., Martins, M., Beerwerth, R., {et~al.} 2017, ApJ, 849, 5,
  \dodoi{10.3847/1538-4357/aa8fcc}

\bibitem[{Schippers {et~al.}(2020)Schippers, Buhr, Borovik, Holste,
  Perry-Sassmannshausen, Mertens, Reinwardt, Martins, Klumpp, Schubert, Bari,
  Beerwerth, Fritzsche, Ricz, Hellhund, \& M\"{u}ller}]{Schippers2020}
Schippers, S., Buhr, T., Borovik, Jr., A., {et~al.} 2020, XRS, 49, 11,
  \dodoi{10.1002/xrs.3035}

\bibitem[{Schlapp {et~al.}(1995)Schlapp, Trassl, Salzborn, McCullough,
  McLaughlin, \& Gilbody}]{Schlapp1995a}
Schlapp, A., Trassl, R., Salzborn, E., {et~al.} 1995, NIMPB, 98, 525,
  \dodoi{10.1016/0168-583X(95)00180-8}

\bibitem[{Verner {et~al.}(1993)Verner, Yakovlev, Band, \&
  Trzhaskovskaya.}]{Verner1993a}
Verner, D.~A., Yakovlev, D.~G., Band, I.~M., \& Trzhaskovskaya., M.~B. 1993,
  ADNDT, 55, 233, \dodoi{10.1006/adnd.1993.1022}

\bibitem[{Viefhaus {et~al.}(2013)Viefhaus, Scholz, Deinert, Glaser, Ilchen,
  Seltmann, Walter, \& Siewert}]{Viefhaus2013}
Viefhaus, J., Scholz, F., Deinert, S., {et~al.} 2013, NIMPA, 710, 151,
  \dodoi{10.1016/j.nima.2012.10.110}

\bibitem[{Westphal {et~al.}(2019)Westphal, Butterworth, Tomsick, \&
  Gainsforth}]{Westphal2019}
Westphal, A.~J., Butterworth, A.~L., Tomsick, J.~A., \& Gainsforth, Z. 2019,
  ApJ, 872, 66, \dodoi{10.3847/1538-4357/aafb3b}

\bibitem[{Zafar {et~al.}(2019)Zafar, Heintz, Karakas, Lattanzio, \&
  Ahmad}]{Zafar2019}
Zafar, T., Heintz, K.~E., Karakas, A., Lattanzio, J., \& Ahmad, A. 2019, MNRAS,
  490, 2599, \dodoi{10.1093/mnras/stz2827}

\bibitem[{Zhukovska {et~al.}(2018)Zhukovska, Henning, \& Dobbs}]{Zhukovska2018}
Zhukovska, S., Henning, T., \& Dobbs, C. 2018, ApJ, 857, 94,
  \dodoi{10.3847/1538-4357/aab438}

\end{thebibliography}

\end{document}